\documentclass[
preprintnumbers,
 amsmath,amssymb,
 aps,
]{revtex4-2}

\usepackage{graphicx}
\usepackage{dcolumn}
\usepackage{subcaption}
\usepackage{url}
\usepackage[dvipsnames]{xcolor}
\usepackage{bm}
\newcommand\Rey{\mbox{\textit{Re}}}

\begin{document}

\UseRawInputEncoding


\title{Aspect-ratio effect on the wake of a wall-mounted square cylinder immersed in a turbulent boundary layer}

\author{Gerardo Zampino}
 \email{gzampino@kth.se}
\affiliation{FLOW Engineering Mechanics\\ KTH Royal Institute of Technology\\ Stockholm, Sweden} 

\author{Marco Atzori}
\affiliation{Dipartimento di Scienze e Tecnologie Aerospaziali\\
Politecnico di Milano\\
Milano, Italy}

\author{Elias Zea}
\affiliation{MWL, Engineering Mechanics\\ KTH Royal Institute of Technology\\ Stockholm, Sweden}

\author{Evelyn Otero}
\affiliation{CSA, Engineering Mechanics\\ KTH Royal Institute of Technology\\ Stockholm, Sweden}

\author{Ricardo Vinuesa}
\email{rvinuesa@mech.kth.se}
\affiliation{FLOW, Engineering Mechanics\\ KTH Royal Institute of Technology\\ Stockholm, Sweden}

\begin{abstract}
The wake topology developing behind a wall-mounted square cylinder in a turbulent boundary layer has been investigated using a high-resolution large-eddy simulation (LES). The boundary-layer thickness at the obstacle location is fixed, the Reynolds number  {based on the cylinder $h$ and the incoming free-stream velocity $u_\infty$} is 10,000 while the aspect ratio (AR), defined as obstacle height divided by its width, ranges from 1 to 4.
The Reynolds stresses, anisotropy-invariant maps (AIM) and the turbulent kinetic energy (TKE) budget are analyzed to investigate the influence of AR on the wake structures and on the turbulence production and transport. In particular, the transition from a dipole configuration for low AR to a quadrupole wake is extensively discussed and examined.  {The necessity of more data to express this critical AR as a function of the momentum-thickness-based Reynolds number $\Rey_\theta$ is thus highlighted.} As an effect of the AR, the wake is deformed in both streamwise and spanwise directions. This contraction of the wake, attributed to the occurrence of the base vortices for the cases $\mathrm{AR}=3$ and 4, impacts the size of the positive production region that stretches from the roof and the flank of the obstacle to the wake core. The AIMs confirm the wake three-dimensionality and are used to describe the redistribution of the turbulent kinetic energy (TKE) along the three normal directions, in agreement with the literature [A. J. Simonsen and P. Krogstad, Phys. Fluids 17, 088103, (2005)].  {The present analysis on the TKE budget displays a stronger turbulence production for the cases $\mathrm{AR}=3$ and 4, demonstrating the strong influence of the tip and base vortices in generating turbulence at the wall location behind the cylinder.} 
\end{abstract}

\maketitle


\section{Introduction}
The study of turbulent flows in urban areas has gained increased attention from the scientific community for two primary reasons. Firstly, 75\% of the European population lives in highly polluted cities, with 90\% of all urban population exposed to elevated levels of smog and pollutants. Predicting turbulent flows around buildings and modeling pollutant dispersion becomes crucial for achieving Sustainable Development Goal 11 \citep{UN_SDG}, which aims to make cities safe, sustainable, and inclusive.
Secondly, the use of unmanned aerial vehicles (UAVs) for goods delivery in cities requires a thorough examination of turbulence around buildings for optimizing costs, trajectory planning, and mitigating potential collisions with buildings. The pollutant dispersion \citep{Simoens2007,baj2019,kirkil2020,Keshavarzian2021,Lim2022}, heat exchange \citep{Sohankar2019} and the turbulent structures have been extensively studied in literature for urban-like configurations immersed in a turbulent boundary layer \citep{castro2006,Torres2021,Tian2021,Lazpita2022}. Both numerical and experimental studies have been published with the aims of discussing the dependence of the near-wake structures on the cross-section of the wall-mounted cylinders \citep{Kumar2019,Rastan2021}, the Reynolds number \citep{Zhao2021}, the turbulence of the incoming flow \citep{Vinuesa2015} and the boundary-layer thickness 
\citep{Chen2022,ElHassan2015}. 
\citet{Nagib1984} firstly studied the correlation between the wake turbulent structures on the pedestrian comfort, followed by more recent studies \citep{monnier2010,Jansen2013,Monnier2018,Amor2023}. In particular, \citet{Monnier2018} experimentally studied the interaction of wind gusts and large-scale vortical structures at different incidence angles and correlated the arch vortex developing behind the obstacles with the region of high and low turbulent kinetic energy. The increase in the AOI leads to a reduction of coherence between the velocity fluctuations in the streamwise and spanwise directions and this affects the strength of the wind gusts that are correlated with the pedestrian comfort \citep{ahmad2017}. 
In this context, \citet{BitterHanna2003} investigated the transport and diffusion of pollutants in an urban environment across various length scales, ranging from the landscape to the neighborhood. 
Other authors studied the propagation of a passive scalar behind a wall-mounted cylinder \citep{Keshavarzian2021}.

A second category of studies focuses exclusively on the flow dynamics behind wall-mounted cylinders \citep{Wang2009,rastan2017,Tian2021}. Since the initial experimental investigations regarding circular cylinders \citep{etzold1976,sakamoto1983,Kawamura1984}, many authors agree that the wake structure depends on the aspect ratio (AR), defined as the ratio of the cylinder height to its diameter \citep{sakamoto1983}.  {The geometrical aspect ratio is thus a critical parameters also for internal flows, for instance it affects the properties of secondary flows emerging at the corners of non-circular ducts \citep{vinuesa2014,samanta2015}.}
For AR below a critical value $\mathrm{AR}_\mathrm{c}$, the vortex shedding behind the wall-mounted cylinder is symmetric and it is characterized by an arch-type vortex as observed by \citet{sakamoto1983} and \citet{Kawamura1984}.  {The vortex topology plays a crucial role in influencing the production and distribution of wake turbulence \citep{Atzori2023}. A comprehensive understanding of this phenomenon is thus essential for optimizing drone trajectories and avoiding regions characterized by high turbulence-production levels, which may negatively impact drone-flight performance. However, the} critical value of $\mathrm{AR}_\mathrm{c}$ is not a universally agreed-upon definition and depends on many factors, such as the turbulence characteristics of the incoming flow \citep{Vinuesa2015} and the boundary-layer thickness \citep{Hosseini2013,ElHassan2015,Chen2022}. The critical value is estimated to be between $2$ and $6$ \citep{sakamoto1983,Kawamura1984,pattenden2005,sumner2004,Wang2009} but an analytical expression is still missing in the literature. Beyond the critical value, the wake loses its symmetry, and vortex shedding becomes antisymmetric, leading to the well-known K{\'a}rm{\'a}n vortex shedding. For both symmetric and antisymmetric vortices, a horseshoe vortex has been identified as an elongated vortical structure that occurs at the junction between the cylinder and the bottom surface due to the adverse pressure gradient induced by the obstacle blockage \citep{Hussein1996,sumner2004,Sahin2007}. This horseshoe vortex envelops the flanks of the cylinder and slightly affects the outer portion of the wake. 
Although the aforementioned authors primarily focused on the analysis of circular cylinders, the effect of the cylinder shape on the wake characteristics has been scarcely investigated. 
For example, \citet{Kumar2019} conducted a comparative study on the wake structures of three wall-mounted cylinders with different cross sections (circular, triangular, and square) using direct numerical simulations (DNS) at varying Reynolds numbers and as a function of the shear intensity. Despite the fact that AR was the same for all cases, the transition from symmetric to asymmetric vortex shedding occurred first for the circular cylinder and then for the square and triangular cylinders with changing Reynolds numbers. A noteworthy study in this regard is the work conducted by \citet{Keshavarzian2021}, who investigated the effect of the cross-section shape on the dispersion of the pollutant in the wake of cylinders with a  {polygonal cross-section, ranging from a square to a circular section}. \citet{Keshavarzian2021} concluded that the stronger downwash observed for the circular cylinder resulted in higher pollutant concentrations in the wake compared to the square cylinder.  
\citet{McClean2014} first observed the effect of the yaw angle for the incoming flow on the wake structures of a wall-mounted square cylinder with varying AR. \citet{McClean2014} observed that the transition from symmetric to antisymmetric vortex shedding can be forced by increasing the yaw angle. 

 {The present study is motivated by the lack of high-resolution LES investigating the influence of the wall-mounted cylinder aspect ratio on tip, base, and vortex shedding behind a wall-mounted square cylinder. Although some authors, \citet{Zhao2021} among others, have classified the near wake based on AR, their focus was on low Reynolds numbers. For this reason, we conducted a series of high-resolution LESs, with resolutions slightly coarser than those in DNS \citep{Atzori2023}, at relatively high Reynolds numbers and subjected to incoming turbulent boundary layers.}

\subsection{Wake description}
In this framework, it is worth mentioning the work of \citet{Wang2009} for a more comprehensive explanation of wake turbulent structures behind a wall-mounted cylinder.
\citet{Wang2009} proposed an empirical interpretation of the turbulent wake as the synthesis of three structures: tip, base and spanwise vortices. The tip vortices are a couple of counter-rotating vortices developing at the free end of the cylinder roof and producing a strong downwash in the wake. In proximity to the base,  a second pair of counter-rotating vortices develop  with an opposite sense of rotation in respect to the tip vortices. The base vortices are inclined upwards and they produce a strong upwash in the lower portion of the wake. Both tip and base vortices extend in the streamwise direction downstream the cylinder and were used by \citet{Wang2009} to distinguish between a ``dipole", showing only the footprint of the tip vortices, and ``quadrupole" configuration, where both tip and base vortices are observed. However, as further discussed by \citet{Zhang2017}, the base vortices weaken downstream and the ``quadrupole" in the near-wake region changes into a ``dipole" configuration. The spanwise vortices are defined as the turbulent structures separating from the flank of the cylinders. 

\citet{sakamoto1983} hypothesized that the flow separating from the roof of the cylinder creates a turbulent filament that bridges the spanwise vortices from the flank of the cylinder. 
 \citet{sumner2004} interpreted the base vortices as the oblique separation of the spanwise vortices. As a consequence, the spanwise vortices are inclined upwards and they are streamwise-developing rolls. Following these assumptions, \citet{Wang2009} speculated that the spanwise vortices at the lateral of the cylinder consist of two vertical turbulent filaments that are linked to each other by a horizontal bridge forming an arch-type vortex. Thus, the spanwise vortices can be symmetrically arranged (symmetric wake) or staggered (antisymmetrical  or K\'{a}rm\'{a}n vortex shedding). The tip and base vortices are the footprint in a vertical section of the arch-type vortex theorized by \citet{sakamoto1983}.
The model from \citet{Wang2009} is an improved version of the model proposed by \citet{Kawamura1984} and \citet{sumner2013}, where tip, base and spanwise vortices were three independent flow structures. 

\citet{bourgeois2011} proposed a fully connected interpretation of the wake. Two vertical principal cores occur at the lateral edge of the buildings and both sides are connected by a horizontal ``connector strand" \citep{bourgeois2011,bourgeois2012}. The discernible case between a ``dipole" configuration or ``quadrupole" wake is the presence of two looped or half-looped vortices that oscillate behind the obstacle. 
Recently, \citet{Zhang2017} studied the effect of the Reynolds number on the wake structure for a finite wall-mounted square cylinder with $\mathrm{AR}=4$  and low Reynolds numbers. \citet{Zhang2017} first observed a third wake configuration known as ``Six-Vortices Type", where six poles develop behind the obstacle. Another model is the hairpin vortex model by \citet{Dousset2010}, which limits the description of the instantaneous wake. 
According to this model, for both symmetric and antisymmetric wake, the ``legs" at the sides of the building form a ``C" shape in any horizontal plane in the wake of the building. Moving downstream, the two ends of the ``C" shape are deflected upstream forming the ``Reverse-C" spanwise vortices \citep{Zhang2017}. These latter turbulent structures transit to the hairpin vortices due to the fragmentation of large pieces of vortices because of the interaction between the boundary layer, the shear layer, and the downwash produced by the free end of the cylinder \citep{Zhang2017}.  
Finally, a conceptual model has been proposed by \citet{Rastan2021} to further explain the effect of the cross-section aspect ratio (CR) on the flow separation and wake using both numerical and experimental setups. Here, CR is defined as the ratio between the cylinder longitudinal length and its width. \citet{Rastan2021} observed that CR affects more significantly the wake structures than AR and for CR=1 the wake is characterized by two streamwise vortices originating from side edges forming a dipole configuration. For increasing CR, the flow reattaches on the top surface. This creates a separation bubble that merges with the streamwise vortices and forms streamwise elongated structures known as the side vortex. Above the roof of the cylinder, a second pair of weaker tip vortices occurs but they rapidly degrade.  The reattachment of the flow breaks the three-dimensionality of the flow and becomes more two-dimensional. Given these considerations, \citet{Rastan2021} proposed that, in the absence of reattachment (low CR), the near wake displays an arch-shaped coherent structure with staggered legs at the flanks. For CR higher than 2, the reattachment on both top and side surfaces alters the wake characteristics, eliminating large-scale vortices that weave and intersect within the wake. 

Although Reynolds-averaged Navier--Stokes (RANS) simulations have been employed to study these structures, they seem inadequate to fully characterize the physics of the wake generation. Therefore, DNS \citep{Saeedi2014, DiazDaniel2017} and large-eddy simulations (LES) are widely employed \citep{Tian2021,Zhao2021}. \citet{yauwenas2019} combined experiments and LES of a wall-mounted square cylinder in a turbulent boundary layer. Their study explores the parametric change of the wake structures as a function of the AR. By considering their results with those of other authors such as \citet{Wang2009,bourgeois2011,Hosseini2013},  \citet{yauwenas2019} developed a new diagram of the transition from a dipole to a quadrupole wake as a function of two parameters: the aspect ratio and the boundary-layer thickness. \citet{yauwenas2019} highlighted the existence of a transition region where both wake configurations are equally possible.


The methodology and numerical setup are extensively discussed and motivated in Section \ref{chap:Comp_setup}. The influence of the aspect ratio (AR) on the wake turbulence is detailed in Section \ref{chap:Effect_AR}, supported by the analysis of anisotropy-invariant maps in Section \ref{chap:Anisotropy}, and Reynolds stresses  and the turbulent kinetic energy (TKE) budget are discussed in Section \ref{chap:Reynolds} and Section \ref{chap:TKE}, respectively. Conclusions and final considerations are presented in Section \ref{chap:Conclusion}.

\section{Computational setup} \label{chap:Comp_setup}

 A wall-mounted square cylinder immersed in an incompressible turbulent boundary layer is investigated using high-resolution LES.
The instantaneous streamwise $u$, vertical $v$ and spanwise $w$ velocities along the $x$, $y$, and $z$ axes, respectively, are Reynolds averaged and decomposed into a mean quantity in time $\overline{\textbf{u}}$ and a turbulent fluctuation $\textbf{u}'$. The Reynolds stresses are hence denoted by $\overline{u_i'u_j'}$.
The governing equations are dimensionless and scaled by the height of the cylinder $h$ and the free-stream velocity $u_{\infty}$ of the incoming flow. Given these definitions, the Reynolds number is $\Rey_h= u_{\infty} h/ \nu$, where $\nu$ is the kinematic viscosity of the flow, and the friction Reynolds number is defined as $\Rey_{\tau}= u_\tau h /\nu$, where $u_\tau=\sqrt{\tau_w/\rho}$ and $\tau_w$ the wall-shear stress. The momentum-thickness-based Reynolds number is $\Rey_\theta=u_\infty \theta /\nu$, where the momentum thickness is $\theta$.

\begin{figure}
    \centering
    \includegraphics[width=0.45\textwidth]{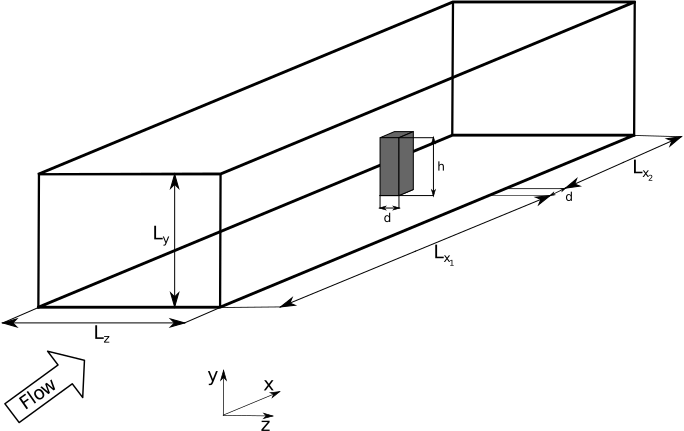}
    \caption{Sketch of the domain and the obstacle. }
    \label{fig:disegno}
\end{figure}

The high-resolution LES are carried out with the open-source code Nek5000, which was developed by \citet{fischer2008} using the spectral-element method (SEM) as described by \citet{patera1984}. The governing equations are discretized using the $P_n-P_{n-2}$  formulation of the Galerkin projection method, where $n$ represents the maximum polynomial order of the trial function for velocity, while $n-2$ corresponds to the polynomial order for the pressure trial function. The total number of elements is between 210,000 to 240,000 for the cases under studies. For each element, a {7th-order} polynomial following Gauss--Lobatto--Legendre (GLL) quadrature has been considered. The resolution is increased by reducing the size of the elements close to the wall. The mesh follows the criteria proposed by \citet{Negi2018} for the same high-resolution LES implementation. In agreement with the simulations carried out by \citet{Atzori2023}, the near-wall resolution follows the guidelines $\Delta x^+ < 18$, $\Delta y^+<0.5$, and $\Delta z^+<9$.
Moving away from the wall, the mesh resolution satisfies the condition $\Delta/\eta < 9$, where $\Delta=(\Delta x \Delta y\Delta z)^{1/3}$, $\eta=(\nu^3/\epsilon)^{1/4}$ is the Kolmogorov scale and $\epsilon$ is the local isotropic dissipation. The high-resolution LES is based on the time relaxation described by \citet{Negi2018} for the approximate deconvolution model. The governing equations include a forcing term defined as the following high-passed filtered velocity:  
\begin{equation}
    \mathcal{H}(\widetilde{u}_i)=\chi \displaystyle \sum_{k=0}^{N} \gamma_k \alpha_k P_k,
\end{equation}
where $\tilde{u}$ is the filtered velocity, $\chi$ is the filter weight, while $P_k$ and $\alpha_k$ denote the base polynomials and the corresponding coefficients, respectively. The filter transfer function is:
\begin{equation}
    \gamma_k= \left\{ 
    \begin{split}
        0 \quad \quad \quad \quad & k_s \leq k_{s_c}, \\
        \left(\frac{k_s-k_{s_c}}{N-k_{s_c}}\right)^2 \quad & k_{s} > k_{s_c},
    \end{split} \right.
\end{equation}
where $k_{s}$ is the wave number and $k_{s_c}$ is the cut-off wave number. The forcing term defined in this way is purely dissipative. The Navier--Stokes equations for the filtered velocity $\widetilde{u}_j$ then become:
\begin{align}
    \label{eq:governing_vel}
    \frac{\partial \widetilde{u}_i}{\partial t}  + \widetilde{u}_j \frac{\partial \widetilde{u}_i}{\partial x_j} &= -\frac{\partial p}{\partial x_i} + \frac{1}{\Rey_h} \frac{\partial^2 \widetilde{u}_i}{\partial x_j \partial x_j} - \mathcal{H} (\widetilde{u}_i),\\
    \frac{\partial \widetilde{u}_i}{\partial x_i}&=0,
\end{align}
Following the resolution criteria reported above, this methodology has been shown to yield excellent agreement for the mean velocity, the Reynolds-stress tensor and the turbulent-kinetic-energy budget. 
The domain size is $L_{x_1}=10h, L_{x_2}=6h, L_y=3h$ and $L_z=4h$ in the $x$, $y$, $z$ directions, respectively (see sketch in Figure \ref{fig:disegno}).  

At the inlet, an incoming Blasius boundary layer profile with a displacement-thickness-based Reynolds number $\Rey_{\delta*}=450$ is imposed.  The transition from laminar to turbulent flow is forced by a numerical tripping consisting of a numerical random volume forcing to model the presence of an experimental trip. The tripping forcing is applied along a transversal line at $x=-9$. The position and the parameters of the tripping have been calibrated to enable fully-developed turbulent boundary layer before reaching the obstacle.  For more details about the tripping function, see \citet{Hosseini2016}. At the top boundary we prescribed a uniform freestream velocity and an outflow, leading to a zero-pressure-gradient (ZPG) boundary layer. In the spanwise direction we impose periodic boundary conditions, and at the outflow we set the stabilized condition by \citet{dong2014}.

In the present analysis we report the results for a wall-mounted square cylinder in a turbulent boundary layer with $\Rey_h=10,000$, corresponding to $\Rey_{\tau} \approx 180$ and $\Rey_\theta \approx 460$  at the location of the obstacle (see \citet{Atzori2023} for additional details).
In the present paper, four main configurations are studied at different aspect ratios $\mathrm{AR}=1,2,3$, and 4. The cases will be denotes as ARx, where x is the corresponding value of the aspect ratio.

\section{Effect of AR on the wake structure} \label{chap:Effect_AR}
\begin{figure*}
    \centering
    \includegraphics[width=0.9\textwidth]{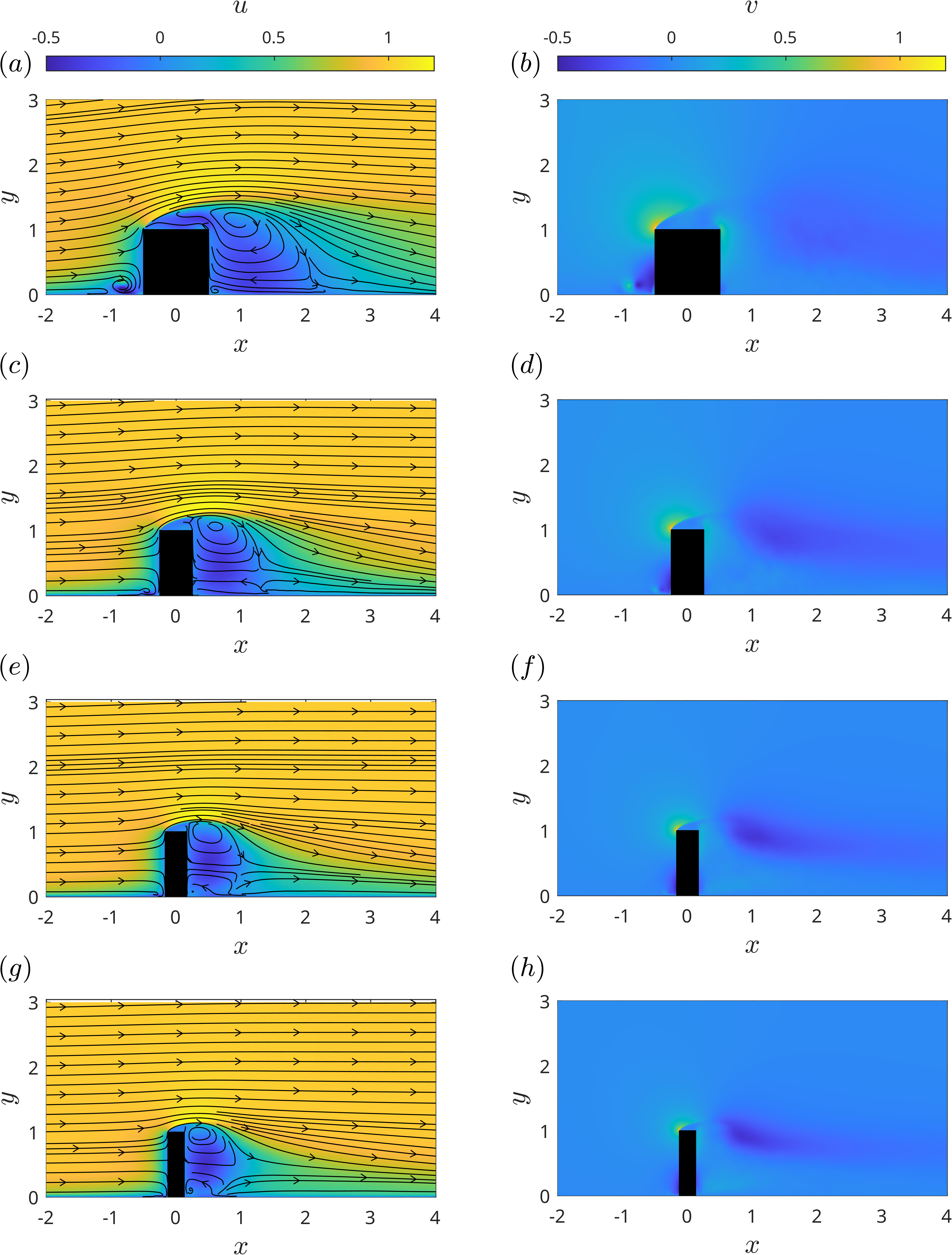}
    \caption{Mean streamwise (left) and vertical (right) velocity components for the cases AR1 (a,b), AR2 (c,d), AR3 (e,f) and AR4 (g,h). The streamlines are shown on the left panels to help the reader to visualize the recirculation bubbles inside the wake.    }
    \label{fig:vel_z}
\end{figure*}
Although the effect of AR on the near wake is well-known in the literature \citep{sakamoto1983, Wang2009, Dousset2010, rastan2017, Zhao2021}, the majority of previous simulations have been conducted at low Reynolds numbers or using a much lower resolution at a higher Reynolds number. In the present study,   {we aim to evaluate the influence of AR on a fully developed turbulent boundary layer without wall modeling. }  
High-resolution LES is used here to better describe the statistical behavior of the wake and to investigate in detail the various terms of the production, transportation, and distribution of turbulence energy throughout the analysis of the TKE budget.

The mean streamwise and vertical velocity components in the vertical symmetry plane are displayed in Figure \ref{fig:vel_z}. Streamlines have been displayed in the same figure to help the reader with the visualization of the recirculation region behind the obstacle. This figure shows that the more evident effect of AR is on the size of the wake because of the increasing influence of the tip and the base vortices. For the case AR1, the tip vortices are weak, and base vortices do not emerge. Thus, the wake is stronger and shows a single recirculation bubble that occupies the entire wake length. The impingement point, defined by \citet{Zhang2017} as the saddle point where both upwash and downwash coexist, is located at the wall and slightly moves upwards for higher AR.
For increasing AR, the wake shrinks in both $x$ and $y$ due to the stronger tip vortices associated with a downwelling motion (see panels f and h). The recirculation bubble is reduced in size, and a second recirculation zone occurs in the lower wake region. This change in the wake structure can also be described by observing the impingement point that moves from the wall (panel a) to the center of the wake (panels e and g). A similar behavior in the change of the wake configuration has been observed by \citet{Zhao2021}, who carried out a series of simulations of square wall-mounted cylinders in a turbulent boundary layer at increasing $\Rey_h$ while its AR is fixed to 4. For increasing Reynolds number, the wake of the wall-mounted cylinder changes, and the impingement point moves from the center of the wake (for $\Rey_h=100$) to the wall (for $\Rey_h=500$). This suggests that both the turbulence characteristics of the incoming flow and the AR are important. 
For case AR1, the flow detaches at the upper edge and reattaches on the roof, forming a small and weak recirculation region. For AR2--AR4, the flow does not reattach on the roof  {because of the shorter width of the cylinders}. In front of the obstacle, a horseshoe vortex emerges. Its size is bigger for AR1 than AR4, where the position of the horseshoe vortex moves downstream until it becomes not discernible from the base vortices. The horseshoe vortex surrounds the sides of the obstacle, causing a downward flow within the vertical tails on either side of the obstacle \citep{Hussein1996}.

In the right panels of Figure \ref{fig:vel_z}, the vertical velocity is broadly similar for all cases studied  {since the separation region remains constant across all the cases.} The maximum positive velocity occurs at the front edge of the roof where the flow detaches, while the outer part of the wake displays a negative $v$ as a result of the tip vortices (see panels f and h). Due to this negative velocity, the wake captures the high-energetic flow at the outer layer and transports it to the core of the wake. The consequences of this mechanism will be clear when analyzing the production term of the TKE budget (see the section \ref{chap:TKE} for more details). The magnitude of the negative vertical velocity is larger for AR1 (panel b), but its intensity is weaker than that of cases AR3 (panel f) and AR4 (panel h). In addition, for the cases AR3 and AR4, in panels (f) and (g) respectively, a positive velocity is barely visible at the bottom part of the wake; note that this is a clear effect of the base vortices.

\begin{figure}
\begin{subfigure}{0.42\textwidth}
\includegraphics[width=\textwidth]{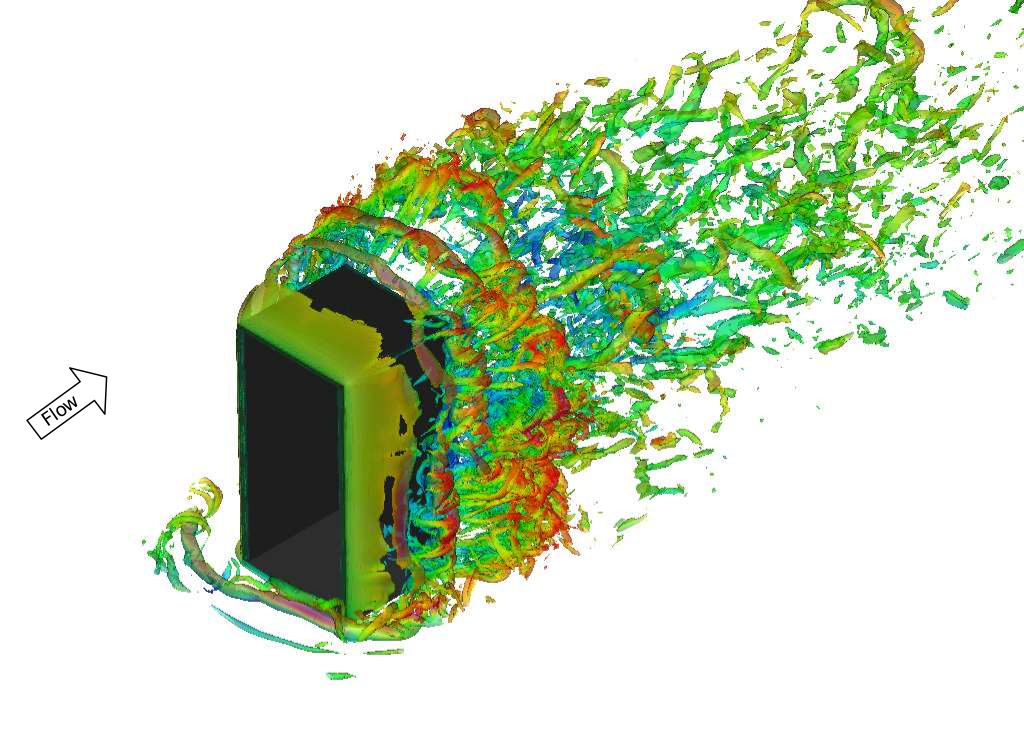}
\quad
\end{subfigure}\hfill
\begin{subfigure}{0.42\textwidth}
\includegraphics[width=\textwidth]{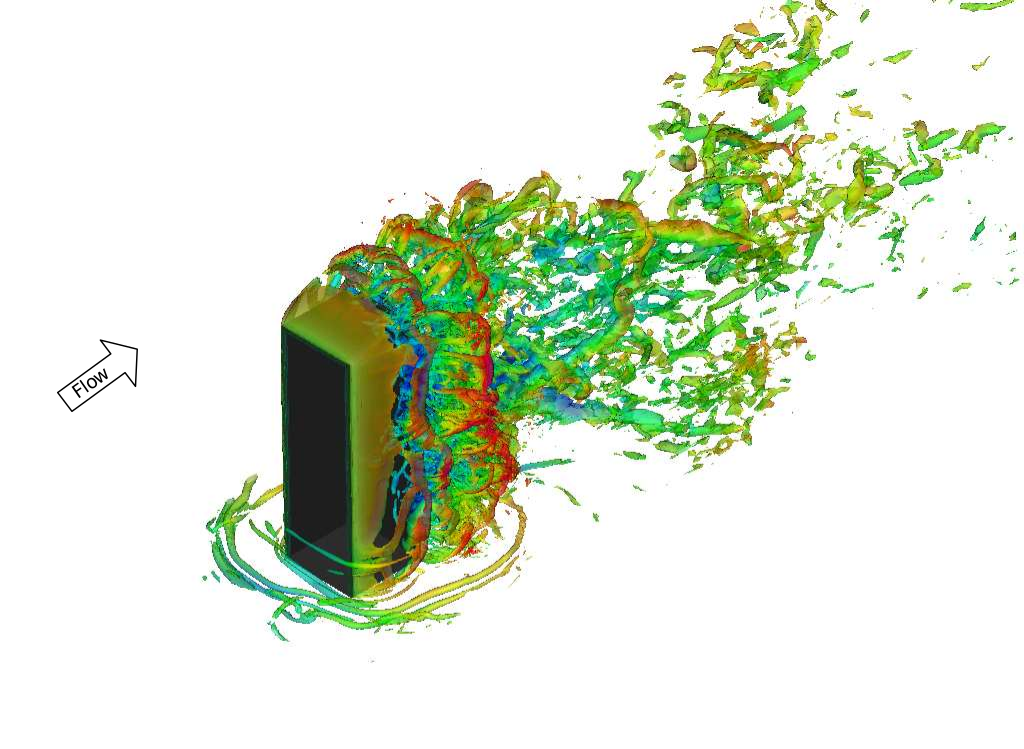}
\quad
\end{subfigure}\par
\begin{subfigure}{0.42\textwidth}
\includegraphics[width=\textwidth]{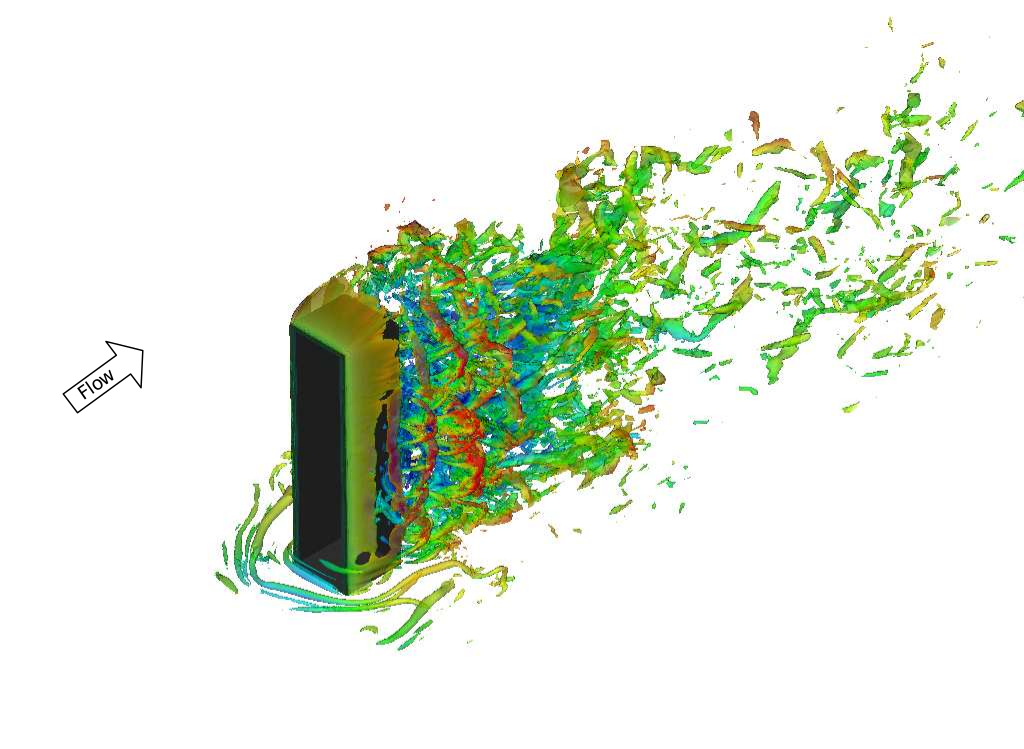}
\end{subfigure}
\caption{ \label{fig:Wake_3D}
Instantaneous vortical motions identified by $\lambda_2=-40$ for the cases AR2 (top), AR3 (centre), AR4 (bottom). The colormap is the instantaneous streamwise velocity $u$, ranging from -1.1 (dark blue) to 1.75 (dark red). The flow direction is indicated by the arrow in each case.}
\end{figure}

The instantaneous vortical motions can be visualized using the $\lambda_2$-criterion \citep{jeong1995} in figure \ref{fig:Wake_3D}.
The wake is symmetric for  {AR1 (not shown) and AR2}, when $\mathrm{AR}< \mathrm{AR}_\mathrm{c}$, and the wake is defined as a dipole. For AR3 and AR4, the wake is antisymmetric (Kármán vortex shedding). In panels (b) and (c), the wake is a quadrupole, and the base vortices are discernible. Downstream of the obstacle, these turbulent structures produce structures resembling hairpin vortices \citep{Dousset2010, Zhang2017} (panel c) agglomerated into a complex turbulent structure.

In this context, \citet{Dousset2010} carried out a series of direct numerical simulations of a wall-mounted square cylinder, immersed in a turbulent flow delimited by a rectangular duct, to investigate the formation of the hairpin vortex for varying Reynolds numbers. 
\citet{Dousset2010} distinguished between a symmetric shedding ($\Rey_d= u_{\infty} d/\nu=200$), where a single row of the hairpin vortices is placed in the streamwise direction, and antisymmetric vortex shedding ($\Rey_d>250$), where the upwash induced by the base vortices forces the hairpin vortices to aggregate in a complex turbulent structure where their parts are barely discernible. This latter wake theory was expanded by \citet{Zhang2017}, who observed that for both the symmetric and antisymmetric configurations, the hairpin vortices are the results of the fragmentation of the Reverse-C spanwise vortices due to the interaction between the downwash, the boundary layer and the shear layer.  Although the hairpin vortices have been already described for low Reynolds numbers, the wake structures observed in Figure \ref{fig:Wake_3D} for higher Reynolds numbers are not related to any of the aforementioned structures. The wake is more chaotic and the only common pattern is the arch-type vortex separating at the cylinder edges. 

\begin{figure}
    \centering
    \includegraphics[width=0.45 \textwidth]{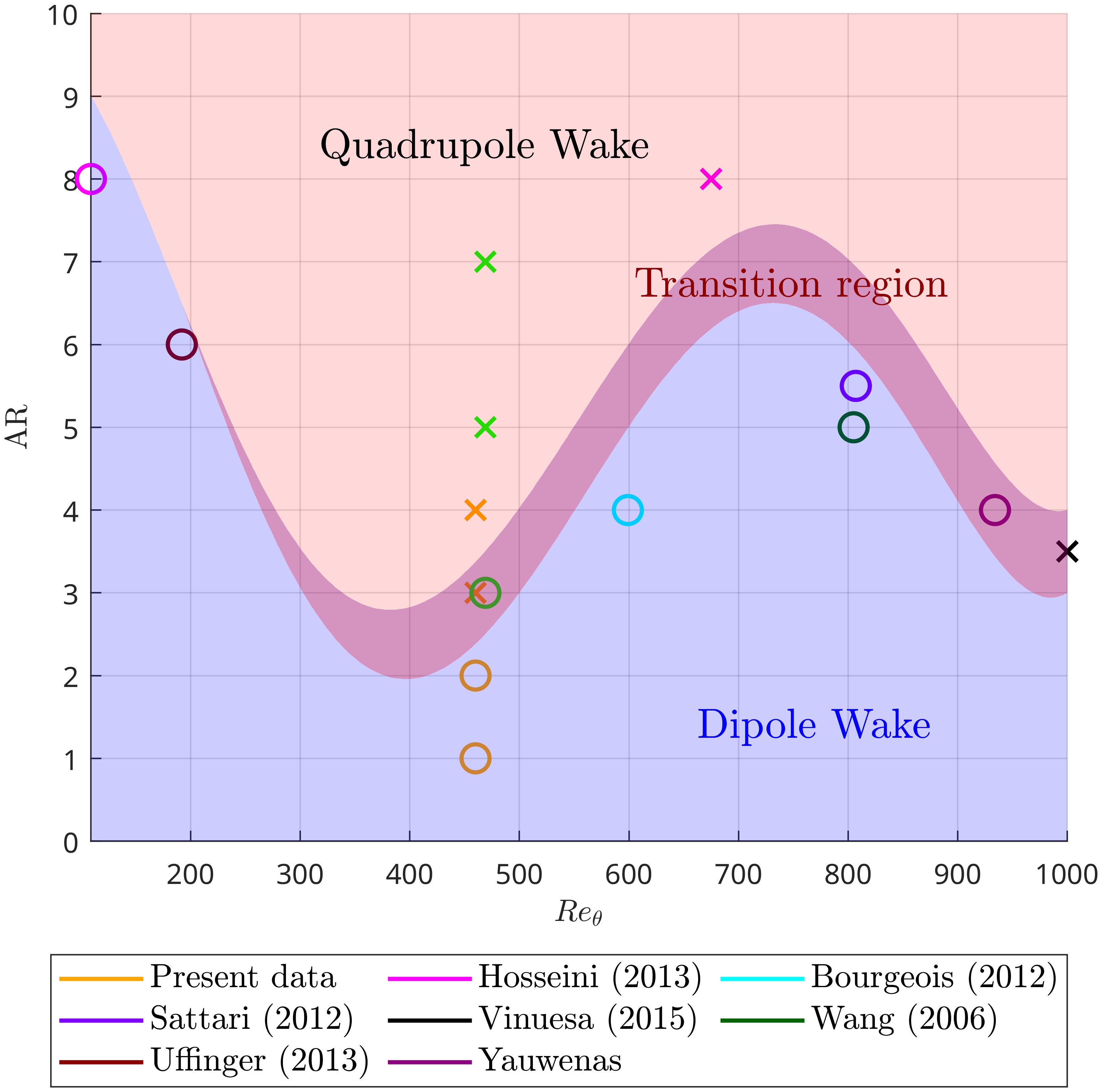}
    \caption{\label{fig:AR_cri} Summary of the wake configuration as a function of  $\mathrm{AR}$ and $\Rey_\theta$ from the present study (orange), \citet{Hosseini2013} (magenta), \citet{bourgeois2012} (cyan),  \citet{sattari2012} (purple), \citet{Vinuesa2015} (black), \citet{Wang2006},\citet{uffinger2013} (dark red), \citet{yauwenas2019} (dark magenta). In the figure, circle markers are used for dipole wakes while cross markers refer to quadrupole wakes.}
\end{figure}

The present analysis suggests a critical aspect ratio ($\mathrm{AR}_\mathrm{c}$) between 2 and 3  {where the wake loses its symmetry and K{\'a}rm{\'a}n-like vortex shedding is observed}. The critical aspect ratio $\mathrm{AR}_\mathrm{c}$ depends on various flow parameters, such as the momentum thickness, the boundary-layer thickness and the turbulence properties of the incoming boundary layer.  {The novelty proposed in this study consists of a qualitative description of the transition between  dipole and quadrupole wakes, say the $\mathrm{AR}_\mathrm{c}$, as a function of $\Rey_{\theta}=u_{\infty} \theta/\nu$, based on the momentum thickness $\theta$. Here, the momentum thickness is preferred because it accounts not only the boundary-layer thickness, as proposed by many authors in literature (see \citet{yauwenas2019} and \citet{cao2022} among others) but also the turbulent intensity and properties of the incoming boundary layer within the available data in the literature \citep{Hosseini2013,bourgeois2012,sattari2012,Vinuesa2015, uffinger2013,Wang2006}}.  {For experiments regarding a turbulent boundary layer developing on a flat plate from \citet{Wang2006},} when $\theta$ is not available, $\Rey_{\theta}$ is calculated by considering a naturally developing turbulent boundary layer. The data of wall-mounted square cylinder immersed in a turbulent boundary layer are reported in Figure \ref{fig:AR_cri} where circular markers denote dipole configurations while cross markers are used for a quadrupole configuration.  {We distinguish a dipole-wake region (blue area) and a quadrupole-wake region (red area) that overlaps in the transition region \citep{yauwenas2019}. \citet{yauwenas2019} observed a transition region where both dipole and quadrupole configurations are possible. Note that \citet{yauwenas2019} considered only the dependence on the boundary-layer thickness, and not the turbulence properties of the flow.  However, it is not enough to characterize the transition between the symmetric and antisymmetric wakes \citep{yauwenas2019}:  as discussed by \citet{Vinuesa2015}, the turbulence characteristics of the incoming flow can alter the wake configurations and this discrepancy is sufficient to explain the difference in the wake configuration between the present data and the results produced by \citet{yauwenas2019}. 
 However, the lack of data within the transition region makes it difficult to establish the exact extension of this area.}
The transition threshold between the dipole and the quadrupole wake configuration changes as a function of $\Rey_{\theta}$ and the AR critical,  {between $\Rey_{\theta}=$460 and 1000, increases with the Reynolds number.}   
The dependence of the $\mathrm{AR}_\mathrm{c}$ on the Reynolds number is well known in the literature and largely investigated by other authors, \citet{Wang2009,Zhang2017} among others. The Figure \ref{fig:AR_cri} summarizes this dependence  {on the $\Rey_\theta$.}   Within a moderate range of $\Rey_\theta$ between 480 and 800, the $\mathrm{AR}_\mathrm{c}$ increases and its value is between 2--6, in agreement with the literature. However, for lower $\Rey_\theta$ between 100 and 400, the critical value rapidly decreases and the dipole wake has been observed experimentally also for very large $AR$ \citep{uffinger2013,Hosseini2013}.  On the other hand, $\mathrm{AR}_\mathrm{c}$ peaks at 5.5 for $\Rey_\theta=809$ and a further increase of $\Rey_\tau$ produces a reduction of $\mathrm{AR}_\mathrm{c}$ that needs to be evaluated.  
 {Although the literature regarding wall-mounted square cylinders immersed in turbulent boundary layers under different conditions is reasonably rich, many authors did not report the momentum thickness, a fact that makes it difficult to provide a more accurate and general expression for $\mathrm{AR}_\mathrm{c}$ at the moment based on Figure \ref{fig:AR_cri}. Additional data will be needed to accurately depict this transition region.}

\subsection{Anisotropy-invariant maps} \label{chap:Anisotropy}
\begin{figure*}
    \centering
    \includegraphics[width=\textwidth]{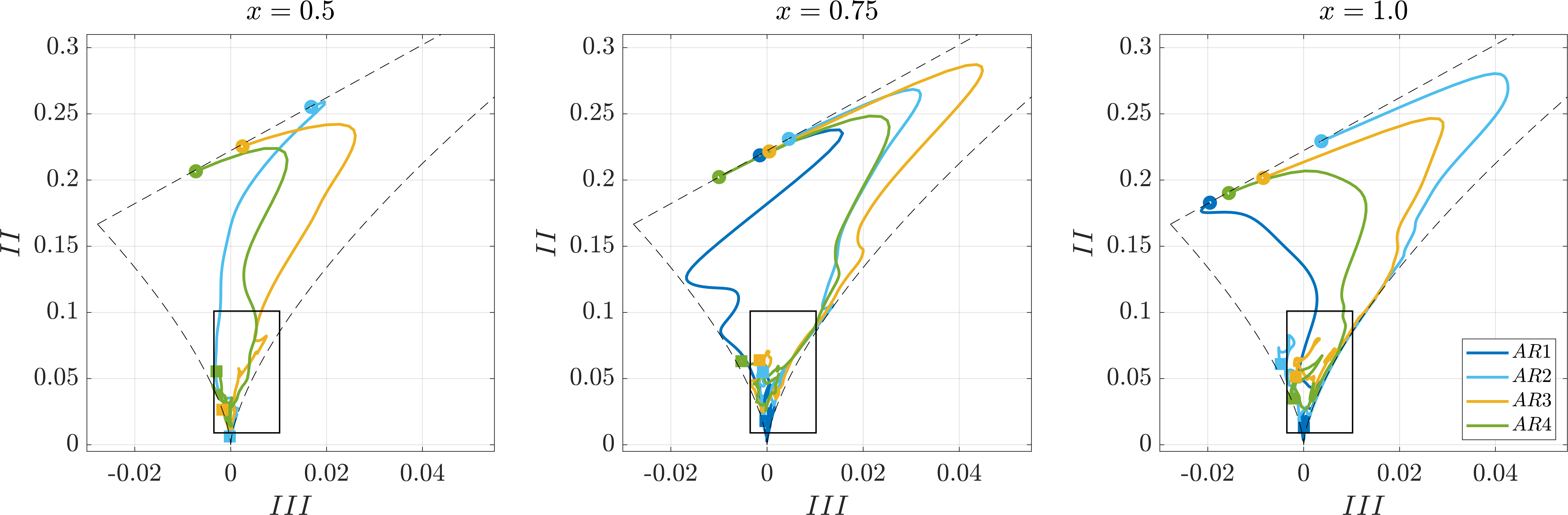}
    \caption{Anisotropy-invariant maps of cases AR1--AR4 (see legend for further details). The velocity profiles are extracted on the symmetry plane ($z=0$) and along the vertical direction between $y=0$ (bottom surface) and $y=1.0$ (obstacle height) and different locations downstream the cylinder at $x=0.5$ (panel a), $x=0.75$ (panel b) and $x=1.0$ (panel b). The black box delimits the region $y$ in [0,0.5]. The black dashed lines are the theoretical limits of the II-III which definitions are reported in equation (\ref{eq:II_III}) \citep{lumley1977}.}
    \label{fig:AIM}
\end{figure*}

The anisotropy-invariant maps (AIM) allow us to visualize the changes in the turbulence distribution across the $x$, $y$, and $z$ directions. The anisotropy tensor was originally introduced by \citet{lumley1977} as: 
\begin{equation}
    a_{ij}=\frac{\overline{u'_i u'_j}}{2k}-\frac{\delta_{ij}}{3}. 
\end{equation} 
\citet{lumley1977} identified three invariants, namely:
\begin{equation}
    I=a_{ii}, \quad \quad \quad II= a_{ij}a_{ji}, \quad \quad \quad III= a_{ij}a_{in} a_{jn}.
    \label{eq:II_III}
\end{equation}
The anisotropy-invariant map (AIM) graphically displays the development of the dynamics of turbulence in the II-III space \citep{lumley1977,Banerjee2007}. The space of the possible II-III values is delimited by the two-component limit (II=2/9+2III) that connects the one-component limit (II=2/3, III=2/9), where the turbulent energy is distributed along a single direction, and the two-component axisymmetric limit (II=1/6, III=-1/36), where the turbulence energy is distributed along two directions. The isotropic limit is given by the conditions II=III=0, which are connected with the 2D-isotropic case by the axisymmetric contraction limit (II=3/2(4/3|III|)$^{2/3}$) and to the 1D limit by the axisymmetric expansion limit (II=3/2(4/3|III|)$^{2/3}$). Using the interpretation proposed by \citet{Simonsen2005}, the nature of the anisotropy can be represented as an ellipsoid whose radius and axis are proportional to the Reynolds stress tensor and its eigenvalue. In agreement with this interpretation, the 1D turbulence corresponds to a line, the two-component limit is represented as a plane elliptical disk that becomes a circular disk for the 2D axisymmetric limit, while a prolate (or oblate) spheroid represents the axisymmetric contraction (or expantion) limit \citep{Simonsen2005}. The limits of the II-III space are reported in panel (a) of Figure \ref{fig:AIM} for the sake of clarification. 

The AIMs for the cases AR1--AR4 at the longitudinal symmetric plane ($z=0$) and for the velocity profiles at $x=0.5$ (panel a), 0.75 (panel b), 1 (panel c) are reported in the same figure. 
The AIMs display the invariants II and III moving from the two-component limits at the base wall (denoted by the circle) to a 3D isotropic limit at the upper region of the wake (denoted by a square marker). This trend is consistent for all positions considered downstream of the obstacle. Except for the first point at the wall, the wake is highly three-dimensional. However, for the cases AR2--AR4, the turbulence distribution follows the axisymmetric expansion limit. For $x=0.75$ (panel b), a different trend is observed for the case AR1 where the curve follows the 2D contraction limit. It is worth noting that the case AR2 is the only case displaying an increase in peak magnitude (panels c and d).  
In the framework of turbulent structures in the urban environment, the effect of the downstream obstacle is more evident when comparing the present analysis with the simulations published by \citet{Atzori2023} for two obstacles in tandem and varying gaps. 
The AIMs from \citet{Atzori2023} for the velocity profiles in the wake of the upfront square cylinder with AR=2 in a two-tandem configuration are included for comparison. Specifically, three configurations have been studied: skimming flow (SF) is shown in panel (a), wake interference (WI) in panel (b), and isolated roughness (IR) in panel (c) at different locations from $x=0.5$ (blue lines) to $x=0.75$ (orange lines) and $x=1$ (yellow lines). These curves are compared with the corresponding maps for an isolated square cylinder (light blue lines).
For IR, the AIM closely aligns with the AR2 case at both $x=0.75$ and $x=1$, while closer to the obstacle ($x=0.5$) the two curves differ.  The difference between the AIMs increases for WI and SF. In this latter case, the turbulence structures observed between the two cylinders and behind the isolated cylinder are very different. In particular, the AR2 case displays a stronger three-dimensionality. The curve at $x=0.5$ is interesting because it follows the asymptotic limit and shows a highly two-dimensional energy distribution. Moving downstream, the turbulence becomes three-dimensional, but the peak of the AIM is lower than the AR2 case for all positions considered.
\begin{figure*}
    \centering
    \includegraphics[width=\textwidth]{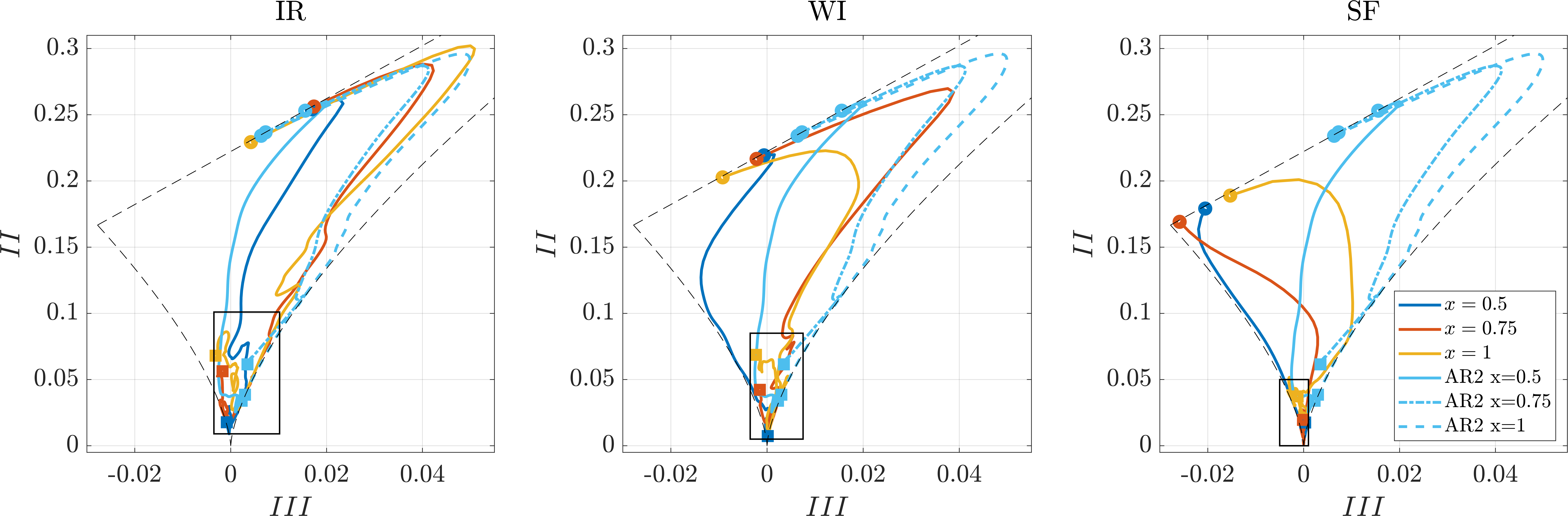}
    \caption{Comparison between the anisotropy invariant maps for two square cylinders in a row with AR=2 and placed at different gaps indicated as skimming flow (SF) in panel (a), Wake interference (WI) in panel (b) and Isolated Roughness (IR) in panel (c). The curves have been obtained at $x=0.5$ (blue lines) to $x=0.75$ (orange lines) and $x=1$ (yellow lines). The data of AR2 (light blue lines) for $x=0.5$ (solid) to $x=0.75$ (dot-dashed) and $x=1$ (dashed) are superimposed. Data extracted from \citet{Atzori2023}.  {The black dashed lines are the theoretical limits of the II-III \citep{lumley1977}}. }
    \label{fig:enter-label}
\end{figure*}

\subsection{Reynolds stresses} \label{chap:Reynolds}
\begin{figure*}
    \centering
    \includegraphics[width=0.95\textwidth]{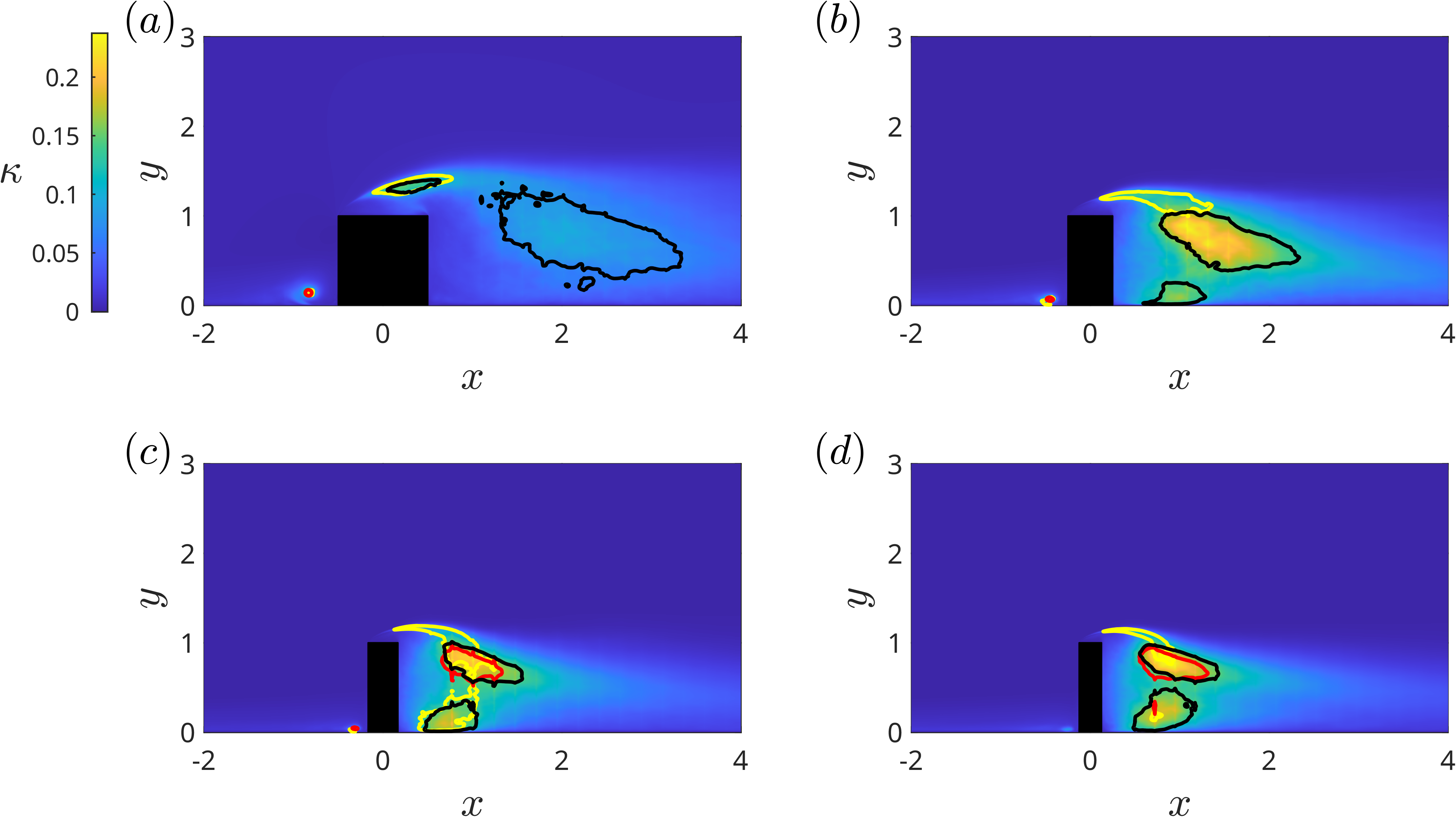}
    \caption{Turbulent kinetic energy $k$  at the symmetry plane $z=0$ for the cases AR1 (panel a), AR2 (panel b), AR3 (panel c) and AR4 (panel d). The contour lines delimit the regions of high $\overline{u'u'}$ (yellow) $\overline{v'v'}$ (red) and $\overline{w'w'}$ (black). The lines correspond to a third of the maximum stress. }
    \label{fig:kappa}
\end{figure*}

From the previous analysis, the wake of a wall-mounted square cylinder is three-dimensional and the turbulent energy is not equally distributed across the three reference directions (see figure \ref{fig:AIM}). The turbulent kinetic energy $k=1/2 (\overline{u'u'}+\overline{v'v'}+\overline{w'w'})$ in the longitudinal symmetric plane ($z=0$) is firstly plotted in Figure \ref{fig:kappa}. The region of high $\overline{u'u'}$, $\overline{v'v'}$ and $\overline{w'w'}$ are also reported in the same figure in order to highlight the correlations between these quantities and the turbulence peak $k$. Upfront the cylinder, the region of higher turbulent kinetic energy corresponds to the footprint of the horseshoe vortex which size and position depends on the AR. For the cases AR1 (panel a) and AR2 (panel b), the horseshoe vortex is larger because of the stronger adverse pressure region upfront the cylinder induced by the building blockage. For increasing AR (panel c and d), the building blockage decreases, the separation line moves towards the upfront surface of the building and the horseshoe vortex reduces in size. For AR3 (panel c) and AR4 (panel d) the horseshoe vortex is negligible and it does not affect the wake strength. 

Apart from the peak of $k$ corresponding to the horseshoe vortex, a second peak emerges above the cylinder roof within the separation region induced by the roof edge. The primary contributor is the streamwise turbulence fluctuation $\overline{u'u'}$ (yellow line), with the other components being negligible or zero. Within the wake, a third, weaker peak is observed in the upper part. This peak aligns with the peak region of $\overline{u'u'}$ (yellow line), $\overline{v'v'}$ (red line), and $\overline{w'w'}$ (black line). For AR3 and AR4, a fourth peak is situated inside the bottom half of the wake, closer to the cylinder base, attributed to the base vortices, with the primary contributors being $\overline{u'u'}$ and $\overline{w'w'}$.

This difference in the turbulence energy can be adduced to the different wake configurations. For AR1 (panel a) and AR2 (panel b) the wake is a dipole, the dominant structure in the wake is the so-called arch-type vortex and the source of the turbulence is located above the cylinder. For the AR3 (panel c) and AR4 (panel d), the peak moves inside the wake region where the most important turbulent structures are the hairpin vortices. 
The wake shrinks in the spanwise direction, the maximum of $k$ tends to occupy the region behind the obstacle and the maximum contribution is given by $\overline{w'w'}$. The streamwise normal stresses peak above the obstacle for AR1 and AR2 where the flow shows a high-shear region. This is probably due to the interaction between the low-momentum flow at the roof of the obstacle and the high-momentum flow in the upper part of the boundary layer. For higher aspect ratios, cases AR3 and AR4, the regions of high $\overline{u'u'}$ above the obstacle are reduced, i.e. the wake is affected by a very strong downwash due to the tip vortices which transports the high-energy flow in the outer part of the boundary layer towards to the low-momentum flow in the wake. This creates a region in the wake where the streamwise momentum $\overline{u'u'}$ is maximum. Inside the wake, $\overline{v'v'}$ is weak and negligible for the case AR1 while it becomes important for higher aspect ratios. The component $\overline{w'w'}$ is equally important inside the wake across all the cases considered in this paper. 
However, the only difference observed between the cases AR1 (low aspect ratio) and AR4 (high aspect ratio) consists of the extended of the peak region. This region is larger for AR1 because of the wider wake while for AR4 the normal stress $\overline{w'w'}$ displays a secondary peak above the obstacle.   

\begin{figure*}
    \centering
    \includegraphics[width=0.9\textwidth]{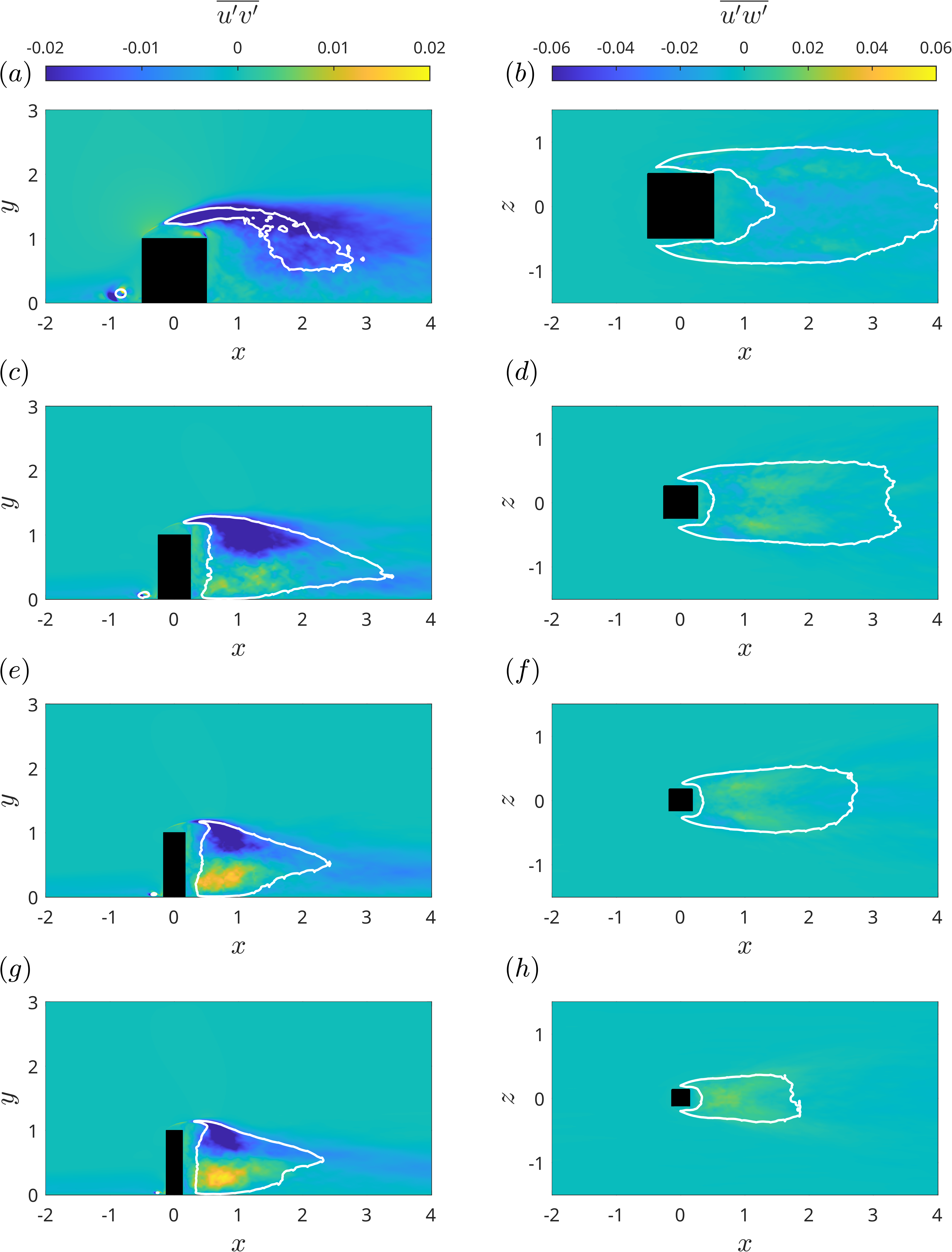}
    \caption{Contour plots of the shear stresses $\overline{u'v'}$ at $z=0$ (left column) and $\overline{u'w'}$ at the horizontal plane $y=0.5$ (right column). The high turbulent kinetic energy $k$ is delimited by the white contour corresponding to 1/3 of the maximum value in the domain for all cases.}
    \label{fig:stress}
\end{figure*}

The shear stresses $\overline{u'v'}$ (left column) and $\overline{u'w'}$ (right column) for the cases from AR1 (top panels) to AR4 (bottom panels) are reported in Figure \ref{fig:stress}. The white contour delimits the region of high TKE. The shear-stress $\overline{u'v'}$ is very negative in the upper part of the wake. This proves the tendency of the tip vortices to transport downwards the high momentum flow behind the obstacle. The positive $\overline{u'v'}$ is displayed at the base of the cylinder and is less intense than the negative region. Thus, the flow is driven towards the wake center but the intensity is quite weak because of the wall. The positive region is stronger for AR4 than AR1 because of the effect of the base vortices that drive the flow from the base away from the wall. Similarly, the horizontal stress $\overline{u'w'}$ displays one negative (left-hand side of the obstacle) and one positive (right-hand side of the obstacle) region that transport the high-energetic flow towards the wake center. The strongest transport term has been predicted for the case AR4.

\subsection{TKE budget} \label{chap:TKE}
The turbulent kinetic energy (TKE) budget is reported to display the effect of AR on the regions of high turbulence production, transport and diffusion. 
Note that the transport equation of the Reynolds-stress tensor is: 
\begin{widetext}
\begin{equation}
\begin{split}
        \frac{\partial }{\partial t} \overline{u_i'u_j'}= -\underbrace{ \overline{u_i'u_k'}\frac{\partial u_j}{\partial x_k}-\overline{u_j'u_k'}\frac{\partial u_i}{\partial x_k}}_{P_{ij}}-\underbrace{2 \nu \overline{\frac{\partial u_i'}{\partial x_k} \frac{\partial u_j'}{\partial x_k}}}_{\epsilon_{ij}} +  \underbrace{\nu \frac{\partial^2}{\partial x_k^2} \overline{u_i'u_j'}}_{D_{ij}}
        -\underbrace{\frac{\partial}{\partial x_k} \overline{u_i'u_j'u_k'}}_{T_{ij}} +\underbrace{u_k  \frac{\partial}{\partial x_k} \overline{u_i'u_j'}}_{C_{ij}}+\\
        -\underbrace{\frac{1}{\rho} \left( \overline{p\frac{\partial u_i'}{\partial x_j}}+ \overline{p \frac{\partial u_j'}{\partial x_i}}\right)}_{\Pi_{ij}^s}-\underbrace{\frac{1}{\rho} \left( \frac{\partial }{\partial x_j} \overline{p u_i'}+\frac{\partial }{\partial x_i} \overline{p u_j'}\right)}_{\Pi_{ij}^t},
        \end{split}
    \label{eq:Reyequation}
\end{equation}
\end{widetext}
where $P_{ij}$ is the production term, $\epsilon_{ij}$ is the pseudo dissipation term, $D_{ij}$ refers to the viscous diffusion, $T_{ij}$ refers to the turbulent transport while $C_{ij}$ is the convection term. The pressure strain and the pressure transport are denoted b $\Pi_{ij}^s$ and $\Pi_{ij}^t$. 
Given the definition of the turbulent kinetic energy (TKE) as $k=\displaystyle \frac{1}{2} \left( \overline{u'u'}+\overline{v'v'}+\overline{w'w'}\right)$, the TKE equation becomes:
\begin{equation}
    \frac{\partial k}{\partial t}=P^k+\epsilon^k+D^k+T^k-\Pi^k-C^k,
    \label{eq:TKEequation}
\end{equation}
where each term is given by the trace of the corresponding tensor on the right-hand side of equation (\ref{eq:Reyequation}). 

\begin{figure*}
    \centering
    \includegraphics[width=0.9\textwidth]{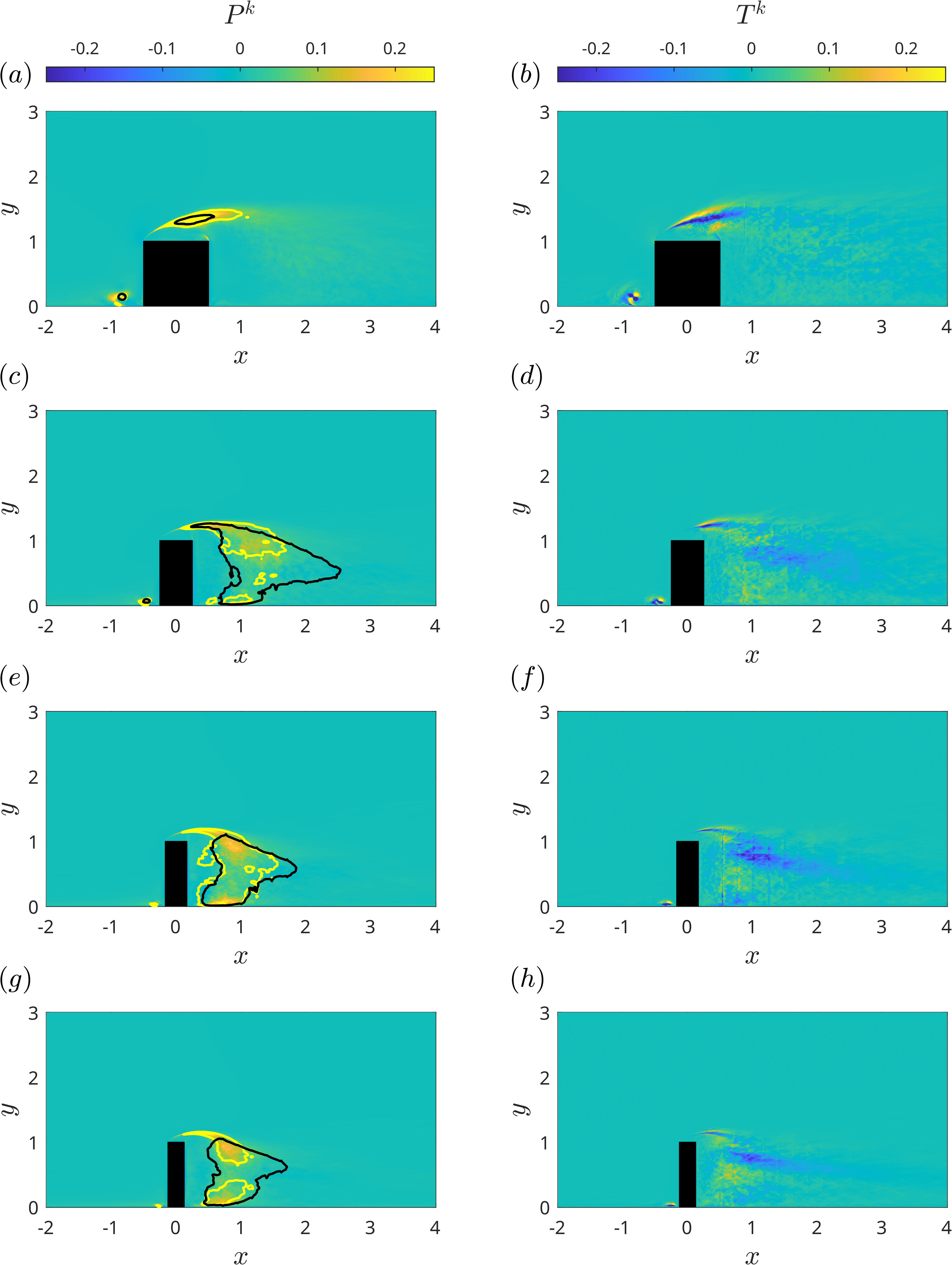}
    \caption{Production $P^{k}$ (left column) and transport $T^{k}$ (right column) terms of the turbulent kinetic energy budget for $z=0$. The cases AR1-AR4 are plotted from the top to the bottom panels.}
    \label{fig:TKE}
\end{figure*}

\begin{figure*}
    \centering
    \includegraphics[width=0.89 \textwidth]{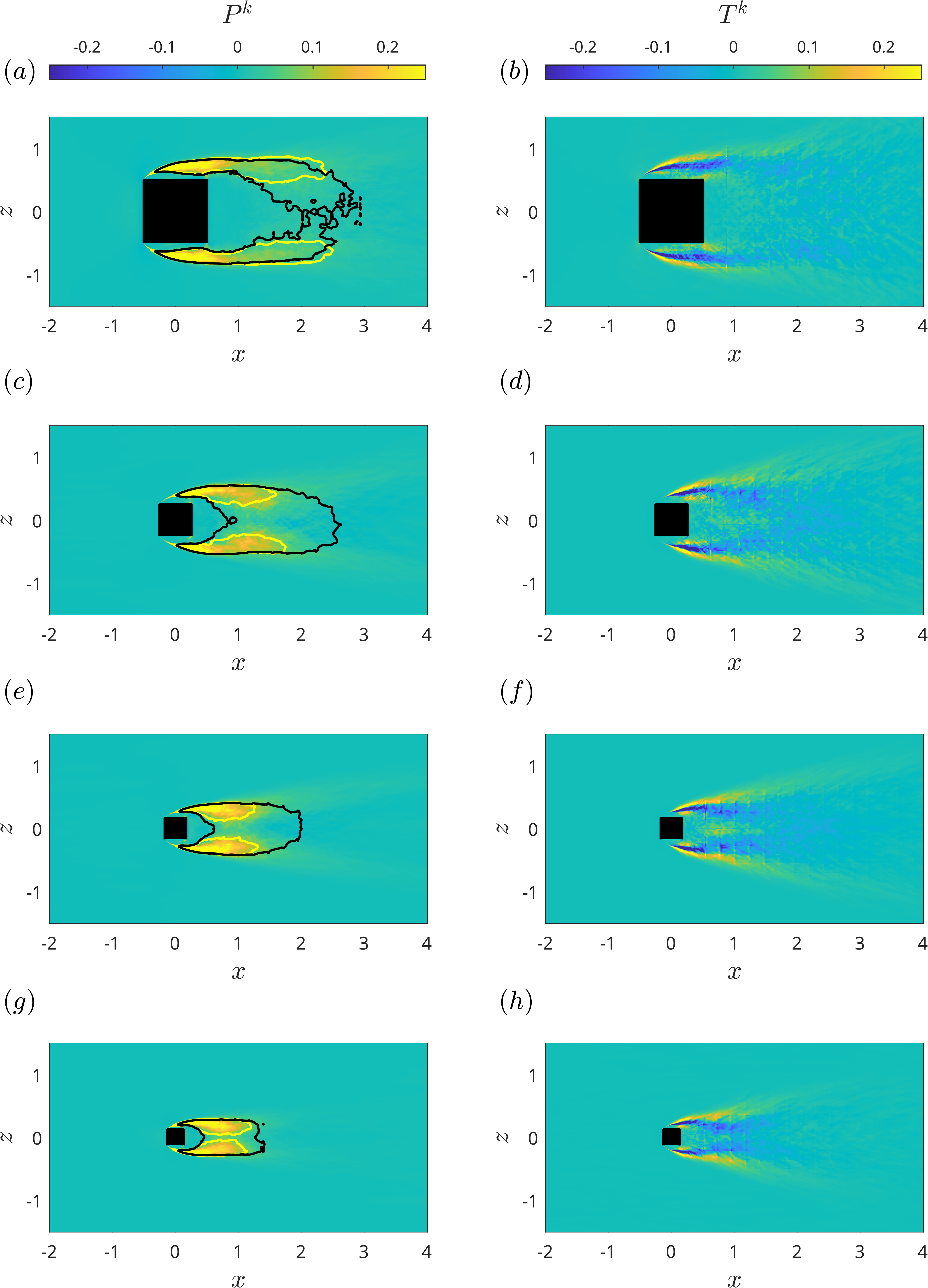}
    \caption{Production $P^{k}$ (left column) and transport $T^{k}$ (right column) terms of the turbulent kinetic energy budget for $y=0.5$. The cases AR1-AR4 are plotted from the upper to the bottom panels.}
    \label{fig:TKE_horizontal}
\end{figure*}

The production $P^k$ and transport $T^k$ terms are shown in Figure \ref{fig:TKE} at the vertical symmetry plane for AR value from 1 (top panels) to 4 (bottom panels). The remaining terms of the TKE budget are not reported for brevity of discussion. The contour lines of the $\overline{u'u'}$ (yellow line) and $k$ (black line) peak regions are also shown. From the figure, both $P^k$ and $T^k$ are negligible in the free stream and assume a nonzero value in proximity to the obstacle and in its wake. In particular, the production term exhibits four strong regions, their size and intensity depend on the aspect ratio (AR): (i) at the roof of the cylinder in the flow separation region, (ii) at the upstream edge of the roof, (iii) upstream of the obstacle in correspondence with the horseshoe vortex, and (iv) at the flank of the cylinders (see Figure \ref{fig:TKE_horizontal}). For the case AR1 (panel a), the peak region of $P^k$ extends to the roof of the cylinder but rapidly degrades in the wake, where it assumes a low positive value. For increasing aspect ratio (left column), the positive production term is distorted, and it is deflected downwards in the wake as a side effect of the tip vortices. Meanwhile, the horseshoe moves towards the upfront face of the cylinder, and its contribution to $P^k$ is confined to a very small region at the wall. For the cases AR3 (panel e) and AR4 (panel g), a weak positive region occurs at the bottom wall due to the base vortices. As observed in Figure \ref{fig:vel_z}, the upper portion of the wake captures the high-energetic flow of the outer boundary layer that is transported to the center of the wake. Similarly, at the bottom surface, due to the occurrence of the base vortices for the cases AR3 and AR4, the flow is pushed upward towards the wake core. As a consequence, the wake becomes the major region of turbulence production. Note that $P^k$ is slightly altered and is negative in front of the cylinder and above the roof of the obstacles. It is worth noting that the maximum of $P^k$ broadly coincides with the peak region of $\overline{u'u'}$ (yellow line).

The transport term $T^k$ is also shown in Figure \ref{fig:TKE}, and it exhibits a negative-positive pattern above the obstacles while it is negligible in the wake. Comparing the left panels, the increasing aspect ratio (AR) produces a deformation of the region of negative $T^k$ that extends into the wake behind the obstacle. The horseshoe vortex is barely visible only for the case AR1, while the transport of turbulence induced by the horseshoe vortex in front of the obstacle is negligible for the remaining cases.
The $P^k$ and $T^k$ in the horizontal plane at $z=0.5$ are finally reported in Figure \ref{fig:TKE_horizontal}. The figure displays a region of high turbulence production at the flank of the obstacles. The turbulence produced within this region is then transported inside the wake by the transport term that shows a positive/negative pattern. For the cases AR2 (panel c) to AR4 (panel g), the positive regions of $P^k$ are deflected from the flank towards the wake center. Similarly, the term $T^k$ has two peak regions at the flank of the obstacles, while for increasing aspect ratio, the wake displays a negative transport term. The case AR1 is the only one that does not display positive production in the wake because it is the only case where the flow reattaches at the roof, the tip vortices are less intense, and their influence on the wake is weak. On the contrary, for the cases AR3 and AR4, the tip and base vortices are stronger and they mainly impact on the turbulence generation in the wake.
From a physical standpoint, the examination of the TKE budget corroborates the idea that the turbulence is initially generated in the flow-separation region and is subsequently transported into the wake. Here, both the tip and base vortices are strongly related to the production of turbulence.

\section{Conclusions} \label{chap:Conclusion}
The present paper aims to describe the influence of the aspect ratio (AR) on the wake of a wall-mounted square cylinder immersed in a turbulent boundary layer with $\text{Re}_h= 10,000$. The computational setup employs high-resolution LES using the open-source software Nek5000. Four main configurations are studied at different AR values (1, 2, 3, and 4) for a better understanding of turbulence behavior in an urban-like environment. The wake is analyzed based on the empirical interpretation by \citet{Wang2009}, which identifies tip, base, and spanwise vortices. In particular, in this paper, we propose a critical value for the emerging base vortices, approximately 2, where the wake exhibits a transition from a dipole configuration (AR1 and AR2) to an antisymmetric K\'arm\'an vortex shedding (AR3 and AR4). The critical value of AR, beyond which an antisymmetric wake is displayed, depends on the momentum thickness and combining the present results with the data published in the literature, {we displayed the dipole, quadrupole and transition regions as a function of $\Rey_\theta$.
 However, more databases are still necessary to better describe the transition between dipole and quadrupole wakes.}
The Reynolds stresses illustrate the anisotropic nature of turbulence, confirmed by the anisotropy-invariant maps (AIM). It is worth noting that in the outer region of the wake, it exhibits strong three-dimensional characteristics except at the bottom wall where the flow is 2D. The turbulent kinetic energy (TKE) budget is finally presented, detailing the contributions of production and transport to turbulence generation in the proximity wake. The production term, influenced by the streamwise transport, is stronger at the roof of the cylinders and in the wake for cases AR3 and AR4, emphasizing the role of tip and base vortices in generating turbulence. The influence of aspect ratio on $P^k$ and $T^k$, especially in regions like the roof, upstream edge, and the sides of the cylinder, highlights that the change in cylinder aspect ratio increases the turbulence  production and transport inside the wake.

The analysis proposed in this study is limited to lower Reynolds numbers than the realistic velocity in urban flows. Although numerical simulations at higher Reynolds numbers are required for a better description of turbulence structures in urban-like configurations, increasing the Reynolds number would impact the computational cost, or make it unfeasible to maintain the same accuracy level, making the high-resolution LES setup unfeasible. 
These results constitute the first step of a more comprehensive analysis in urban-like environments that aims to simulate and optimize the trajectories of drones. The near-wake configuration and the effect of AR on the TKE budget need a more detailed interpretation when considering their impacts on an real-life operating scenario, while the present study describes a simplified configuration. 
Unanswered questions, involving the exploration of additional geometric parameters such as cross-section aspect ratio (CR) or the boundary-layer thickness, and the correlation between the TKE budget with the trajectory optimization of drones will be addressed in future works.

\section*{Data availability statement}
All the data will be available in the following open-access repository after the article is published: \\
\url{https://www.vinuesalab.com/databases/}

\begin{acknowledgments}
This Project has received funding from the European Union’s HORIZON Research and Innovation Programme, project REFMAP, under Grant Agreement number 101096698. The computations were carried out at the supercomputer Dardel at PDC, KTH, and the computer time was provided by the National Academic Infrastructure for Supercomputing in Sweden (NAISS).
\end{acknowledgments}

\bibliographystyle{apsrev4-2.bst}
\bibliography{apssamp}

\end{document}